%% file: main.tex
\documentclass[a4paper]{article}

\usepackage{array, makecell}
\usepackage{multicol}
\usepackage{multirow}
\usepackage{color}
\usepackage{xspace}
\usepackage{booktabs}
\usepackage{hyperref}
\usepackage{graphicx}
\usepackage{pgfplots}
\usepackage{tikz}
\newcommand{\PreserveBackslash}[1]{\let\temp=\\#1\let\\=\temp}
\newcolumntype{C}[1]{>{\PreserveBackslash\centering}p{#1}}
\newcolumntype{R}[1]{>{\PreserveBackslash\raggedleft}p{#1}}
\newcolumntype{L}[1]{>{\PreserveBackslash\raggedright}p{#1}}

\newenvironment{customlegend}[1][]{%
        \begingroup
        \csname pgfplots@init@cleared@structures\endcsname
        \pgfplotsset{#1}%
    }{%
        \csname pgfplots@createlegend\endcsname
        \endgroup
    }%
    \def\addlegendimage{\csname pgfplots@addlegendimage\endcsname}

\usepackage{INTERSPEECH2022}
\newcommand*{\proposed}{CMCD\xspace} 
\newcommand*{\matchingloss}{monotonic matching loss\xspace} 

\newcommand{\ie}{\textit{i.e.}\xspace}
\newcommand{\eg}{\textit{e.g.}\xspace}

\newcommand{\figref}[1]{Figure~\ref{#1}}
\newcommand{\tabref}[1]{Table~\ref{#1}}
\newcommand{\secref}[1]{Section~\ref{#1}}

\newcommand{\equref}[1]{Equation~(\ref{#1})}

\title{Learning Audio-Text Agreement for Open-vocabulary Keyword Spotting}
\name{Hyeon-Kyeong Shin$^{1,2}$ \quad Hyewon Han$^2$ \quad Doyeon Kim$^2$ \quad Soo-Whan Chung$^1$ \quad Hong-Goo Kang$^2$}
\address{
    $^1$Naver Corporation, South Korea\\ $^2$Dept. of Electrical and Electronic Engineering, Yonsei University, South Korea
    }

\email{hkshin@dsp.yonsei.ac.kr}
\begin{document}

\maketitle
\begin{abstract}
In this paper, we propose a novel end-to-end user-defined keyword spotting method that utilizes linguistically corresponding patterns between speech and text sequences. 
Unlike previous approaches requiring speech keyword enrollment, our method compares input queries with an enrolled text keyword sequence.
To place the audio and text representations within a common latent space, we adopt an attention-based cross-modal matching approach that is trained in an end-to-end manner with monotonic matching loss and keyword classification loss.
We also utilize a de-noising loss for the acoustic embedding network to improve robustness in noisy environments.
Additionally, we introduce the LibriPhrase dataset, a new short-phrase dataset based on LibriSpeech for efficiently training keyword spotting models.
Our proposed method achieves competitive results on various evaluation sets compared to other single-modal and cross-modal baselines.

\end{abstract}

\input{0_introduction}

\input{1_background}

\input{2_proposed}

\input{3_Experiment}

\input{4_Conclusion}

\vfill
\pagebreak

\bibliographystyle{IEEEtran}

\bibliography{longstrings,mybib}

\end{document}

%% file: 0_introduction.tex
\section{Introduction}
Keyword spotting (KWS) is the task of identifying enrolled keywords within spoken utterances. 
It is highly challenging because it involves not only detecting keywords accurately, but also rejecting other words reliably.
KWS systems have significantly improved due to the advent of deep learning algorithms, which have demonstrated superior performance in speech analysis and detection.
There are two major approaches toward deep learning-based KWS: one is a keyword-filler strategy that detects phonetic sequences on input streams~\cite{chen2014small, sainath2015convolutional, chen2019small}, and the other is a query-by-example~(QbyE) method~\cite{chen2015query, lugosch2018donut, huang2021query} that matches input queries to enrolled examples.

With increasing demand of user-friendly services in smart speakers and assistants, it has become more important to use user-defined keywords for KWS.
However, detecting user-defined keywords is a more challenging task because it involves dealing with arbitrary phrases which may not have been included in their training data.
Most of user-defined keyword spotting~(UDKWS) methods are designed with QbyE approaches, where the methods detect similarities between a given audio query and pre-enrolled audio examples~\cite{chen2015query, lugosch2018donut, huang2021query}.
However, since QbyE approaches require voice keyword enrollment, their performance highly depends on the consistency of the audio recording session. 
In general, it is not easy to maintain this consistency, so it is difficult to generalize QbyE approaches for a wide variety of users and environments. 
For example, each user has different vocal characteristics. 
And they may enroll the keyword in various recording environments, where input speech can easily be distorted by environmental conditions and background noise.

To address this problem, we introduce the idea of cross-modal approaches with text keyword enrollment instead of speech signals.
We propose an end-to-end UDKWS strategy that utilizes matching criteria between text embeddings and audio embedding queries.
The text modality has only linguistic representations, regardless of users' voices or recording environments.
Therefore, this cross-modal approach can achieve compliant performance results without keyword enrollment issues.
We build a cross-modal correspondence detector~(CMCD), using several sub-modules.
We design two-stream networks to reliably embed linguistic representations of speech and text sequences within a common latent space.
Since the audio-text joint latent space places linguistically similar embeddings close to each other~\cite{sacchi2019open, yusuf2021end}, it is possible to distinguish keywords from other speech inputs.
Based on these representations, our proposed method decides whether the input speech contains a keyword or not, by using a cross-attention mechanism~\cite{vaswani2017attention, momeni2020seeing, lee2021looking}.

We additionally propose a \matchingloss and a de-noising loss to serve as training criteria for our approach.
The \matchingloss enforces correspondence between modalities on the affinity matrix to build robust embeddings, and the de-noising loss helps induce consistent audio representations even if there is noise within the input speech.
We also replace the commonly used binary classification loss with focal loss~\cite{lin2017focal} for more effective learning in imbalanced training sets.
Finally, we introduce the \textit{LibriPhrase} dataset, which contains short phrases (at most 4 words) from the LibriSpeech dataset~\cite{panayotov2015librispeech}.
Since keywords tend to be composed of relatively short sequences of words, LibriPhrase provides more efficient training compared to using a long speech corpus on the open-vocabulary keyword spotting and keyphrase detection tasks.

%% file: 1_background.tex
\begin{figure*}[ht]
    \centering
    \begin{center}
    \includegraphics[clip, width=2\columnwidth]{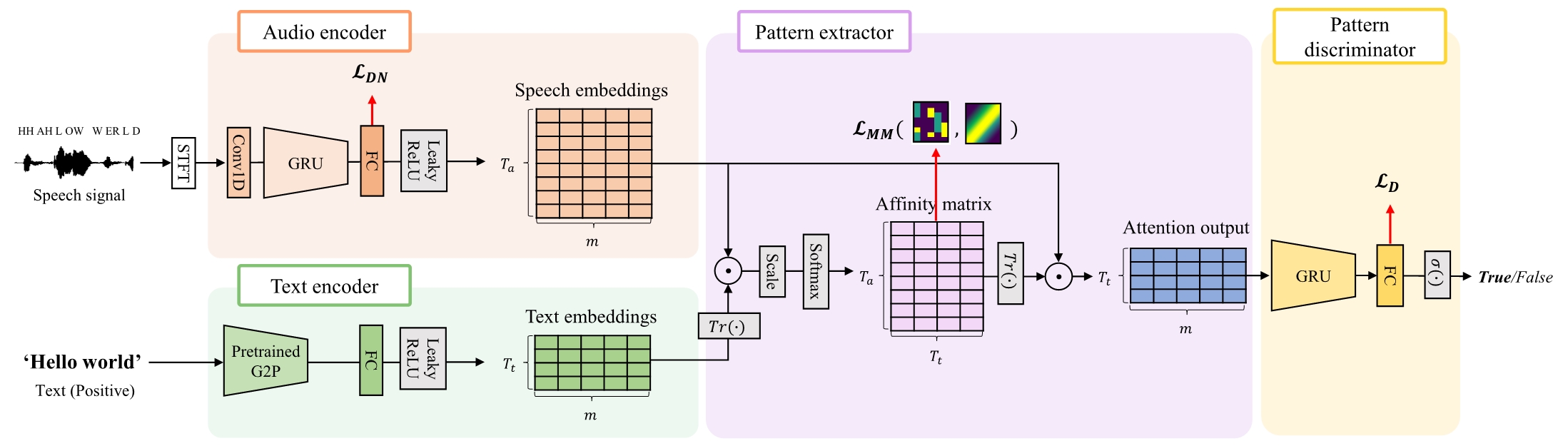}
    \end{center}
    \vspace{-0.5cm}
    \caption{Overall architecture of the proposed model, \proposed. Red arrows indicate each training loss.}
    \vspace{-5pt}
    \label{fig:network_architecture}
\end{figure*}

\vspace{-5pt}
\section{User-defined Keyword Spotting}
\label{sec:related}

In this section, we explain several baseline models that use single-modal or cross-modal approaches for user-defined keyword spotting (UDKWS).
Most previous works for UDKWS utilize query-by-example (QbyE) methods, which use only audio signals as an input.
QbyE methods enroll reference keyword speech and compare it with new input speech queries.
For example, variable-length audio signals with a fixed-length embedding using an LSTM network is represented in~\cite{chen2015query}.
There have been other similar KWS methods using QbyE approaches~\cite{huang2021query}, but their performance degrades for short phrases or non-word inputs due to their word-level training criteria.
In~\cite{graves2006connectionist}, connectionist temporal classification~(CTC) loss, widely used in speech recognition, has been used to overcome the open-vocabulary problem.
An acoustic network trained with the CTC criterion first estimates phonetic sequences, after which a detector determines whether the input speech contains a keyword or not.

Some studies have explored the feasibility of cross-modal approaches that use not only spoken keywords, but also corresponding information from other modalities, \eg, video and text.
In~\cite{sacchi2019open}, an audio-text keyword spotting approach is proposed by computing both audio and phonetic embeddings independently, then mapping them to word-level representations with triplet loss to extract linguistic information and place them in the same latent space.
Also,~\cite{lee2021looking} showed that the word-level embeddings extracted from the enrolled text query are used to compute the probability by multiplying with the frame-wise audio embedding.

Our approach is different from the aforementioned methods because we consider the similarity of the entire text and audio sequences instead of handling word-level embeddings explicitly.
In~\cite{momeni2020seeing}, an audio-visual keyword spotting approach is proposed, where an audio-text model and a visual-text model are pre-trained and then combined into an integrated audio-visual system during the fine-tuning step.
The audio-visual model computes a similarity map between video and audio embeddings and utilizes the similarity map for the keyword detector.
This approach is similar to ours in the sense that it computes similarity map patterns and utilizes these patterns in the keyword recognition stage.
However, this method is difficult to be used in embedded systems due to the large model size.

%% file: 2_proposed.tex
\section{Proposed Method}
\label{sec:proposed}
In this section, we describe our proposed model, a cross-modal correspondence detector (\proposed).
\proposed consists of three sub-blocks: audio and text encoders, a pattern extraction module, and a pattern discriminator.
The overall architecture is illustrated in~\figref{fig:network_architecture}.
\vspace{-5pt}
\subsection{Architecture}
\vspace{-3pt}

\noindent\textbf{Audio/Text encoders.}
Encoders embed audio and text inputs within a joint latent space.
The audio encoder consists of two 1-D convolution layers with a kernel size of 5 and two gated recurrent units~(GRUs). 
The input for the audio encoder is 40-dimensional mel-filterbank coefficients, extracted every 10ms with a 25ms frame length.
For computational efficiency, the first convolutional layer reduces the number of frames by skipping consecutive frames with stride 2.
The text encoder consists of a pre-trained grapheme-to-phoneme~(G2P) model~\cite{g2pE2019} followed by a fully-connected layer.
It looks up the CMU pronouncing dictionary to cover a large number of English vocabulary words.
In the case of out-of-vocabulary~(OOV) words, it estimates the phoneme sequences using a neural network.
We denote audio and text embeddings by $\mathbf{E}_a\in\mathbb{R}^{ T_a\times m}$ and $\mathbf{E}_t\in\mathbb{R}^{ T_t\times m}$, respectively, where $T_a$, $T_t$ and $m$ are the lengths of audio and text embeddings, and the output embedding dimension which is set to 128 in this work.

\vspace{1pt}\noindent\textbf{Pattern extractor.} 
Motivated by~\cite{vaswani2017attention}, the pattern extractor is based on a cross attention mechanism~\cite{momeni2020seeing, lee2021looking} between acoustic and phonetic representations to obtain temporal correlation patterns.
The acoustic embedding $\mathbf{E}_a$ is fed into the network as a key $K$ and a value $V$, and the text embedding $\mathbf{E}_t$ is used as a query $Q$ for the cross attention:
\vspace{-3pt}
\begin{equation}
    Attn=softmax\bigg(\frac{QK^\mathsf{T}}{\sqrt{d_k}}\bigg)\times V = A(Q, K) \times V.\vspace{-3pt}
\label{eq:attn}
\end{equation}
In~\equref{eq:attn}, the affinity matrix $A$ represents the temporal correlation between the audio and text embeddings.
If the $Q$ and $K$ represent the same linguistic content, the $A$ displays a monotonic pattern; if they represent different content, $A$ displays an obscure pattern.
The output of the pattern extractor is the attention matrix $Attn$, which contains information about the audio and text agreement.

\vspace{1pt}\noindent\textbf{Pattern discriminator.} 
The pattern discriminator decides whether audio and text inputs have the same keyword (positive) or not (negative).
A single GRU layer with $m$ dimension takes the attention matrix as an input, and the output of the last frame is fed into a fully-connected layer with a sigmoid function.

\begin{figure}[t]
\begin{minipage}[t]{.24\linewidth}
  \centering
  \centerline{\includegraphics[width=\columnwidth]{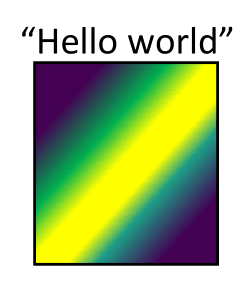}} \vspace{-4pt}
  \centerline{(a)}
\end{minipage}
\begin{minipage}[t]{0.24\linewidth}
  \centering
  \centerline{\includegraphics[width=\columnwidth]{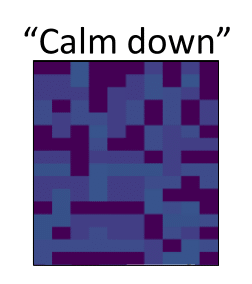}}\vspace{-4pt}
  \centerline{(b)}
\end{minipage}
\begin{minipage}[t]{0.24\linewidth}
  \centering
  \centerline{\includegraphics[width=\columnwidth]{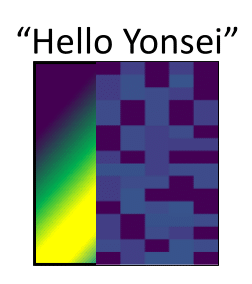}}\vspace{-4pt}
  \centerline{(c)}
\end{minipage}
\begin{minipage}[t]{.24\linewidth}
  \centering
  \centerline{\includegraphics[width=\columnwidth]{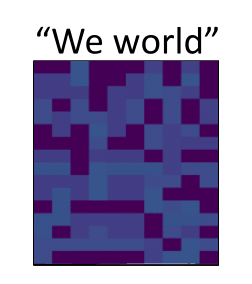}}\vspace{-4pt}
  \centerline{(d)}
\end{minipage}
\vspace{-5pt}
 \caption{Examples of the targets in the \matchingloss when audio keyword is "Hello world"; i) Positive case: (a) Full matching; ii) Negative case: (b) Non-matching, (c) Partial matching (front), (d) Partial matching (back)}
 \vspace{-7pt}
\label{fig:pattern_matching_loss}
\end{figure}

\subsection{Training criterion}
\label{subsec:loss}
\vspace{-3pt}
Our training objective consists of de-noising loss ($\mathcal{L}_{DN}$), \matchingloss ($\mathcal{L}_{MM}$), and detection loss ($\mathcal{L}_{D}$):\vspace{-2pt}
\begin{equation}
\label{eq:total_loss}
   \mathcal{L}_{total} = \lambda_1 \mathcal{L}_{DN} + \lambda_2 \mathcal{L}_{MM} + \mathcal{L}_{D}, \vspace{-2pt}
\end{equation}
where $\mathbf{\lambda_1}$ and $\mathbf{\lambda_2}$ are weighting factors which are set to 0.5, 0.3 in this work.

\vspace{1pt}\noindent\textbf{De-noising loss.}
Motivated by~\cite{qian2015multi, mahto2017vector}, we use a de-noising loss to obtain robust performance even for noisy speech queries.
By constructing the acoustic encoder as a Siamese network, we minimize the mean square error~(MSE) between clean and noisy embeddings as follows:\vspace{-2pt}
\begin{equation}
\label{eq:denoise_loss}
   \mathcal{L}_{DN} = || E_A ^{clean} - E_A ^{noisy}||^2,\vspace{-2pt}
\end{equation}
where $E_{A}^{clean}$ and $E_{A}^{noisy}$ are the acoustic embeddings of clean and noisy speech, respectively.

\vspace{1pt}\noindent\textbf{Monotonic matching loss.}
As noted above, the affinity matrix $A$ in the pattern extractor represents the correspondence between audio and text inputs.
We utilize a monotonic matching loss~(MML) to strengthen the patterns depicted in $A$ by increasing its similarity with manually built patterns.
As illustrated in~\figref{fig:pattern_matching_loss}, we provide three different target affinity patterns $M_i$ depending on the different cases, and minimize the distance between the affinity matrix and a given target pattern during training as shown below:\vspace{-3pt}
\begin{equation}\label{eq:pattern_matching_loss}
  \mathcal{L}_{MM} = ||A - M_i||^2, \quad where \; i\in\{f, n, pf\},
  \vspace{-2pt}
\end{equation}
\begin{equation}\label{eq:full_matching}
M_{f} =\frac{\exp\{-(t_a/T_a - t_t/T_t)^2) / 2g^2\}}{\sum},
\vspace{-2pt}
\end{equation}
\begin{equation}\label{eq:non_matching}
M_{n} =\frac{|X|}{\sum}, \quad where \; X \sim N(\mu, \sigma^2),
\vspace{-2pt}
\end{equation}
\begin{equation}\label{eq:partial_matching}
  M_{pf} = \begin{cases}M_{f},&t_t\leq K\\M_{n},&t_t> K\end{cases},
  \vspace{-2pt}
\end{equation}
where $T_a$, $T_t$ are the lengths of audio and text embeddings, and $g$ is set to 0.2. $K$ is a boundary of matching text.
$X$ is the normal distributed random noise.
$M_i$ is normalized into summation $\Sigma$, to be matched with the range of affinity matrices.

The \textit{full matching} case corresponds to when the input audio and text represent the same keyword; \ie when there is a strong correspondence between the audio and text embeddings.
Here, the pattern on the affinity matrix displays large matching values on the diagonal. 
Inspired by~\cite{tachibana2018efficiently}, we provide an explicit guidance using a Gaussian diagonal pattern $M_{f}$ (\equref{eq:full_matching} and \figref{fig:pattern_matching_loss}(a)).
The \textit{non-matching} case corresponds to when the audio and text represent different keywords; 
the target pattern is a scattered matrix using normalized random noise $M_{n}$~(\equref{eq:non_matching} and~\figref{fig:pattern_matching_loss}(b)).
We also train the affinity matrix on cases where the text is partially matched, the \textit{partial matching} case. 
The target matrix $M_{pf}$ consists of both monotonic and random noise components if the front parts of the sequences match (\figref{fig:pattern_matching_loss}(c)). 
If only the rear parts match, we consider it as the non-matching case $M_{n}$ (\figref{fig:pattern_matching_loss}(d)), because it should be not recognized anymore if the front part is wrong even if the rear part is matched in the keyword spotting.

\vspace{1pt}\noindent\textbf{Detection loss.}
The detection loss $\mathcal{L}_D$ aims to teach the model to detect keywords based on the cross-attention matrix.
Binary cross-entropy~(BCE) loss is commonly used for this objective.
We use BCE loss in the early stages of training, and replace it with focal loss~\cite{lin2017focal} in later stages.
Focal loss shows impressive results when the training data are not well structured (\eg imbalanced dataset~\cite{lin2017focal} or miscalibration~\cite{mukhoti2020calibrating}).
Since we train our network using a general corpus and the number of positive pairs is much smaller than the number of negative pairs, we find that it is advantageous to use the focal loss for training efficiency.

%% file: 3_Experiment.tex
\section{Experiments}
\label{sec:exp}
\subsection{Experimental settings}
\begin{table}
\centering
\caption{Examples of LibriPhrase}
\vspace{-4pt}
\begin{tabular}{|c|l|l|}
\Xhline{1pt}
\bf{Anchor} & \bf{Easy negatives} & \bf{Hard negatives}\\
\Xhline{1pt}
\multirow{3}{*}{friend}
& \small guard & \small  frind \\
& \small comfort & \small rend \\
& \small superior & \small trend \\
\hline
\multirow{3}{*}{the river}
& \small every morning & \small the giver \\
& \small town with & \small the liver \\
& \small not occurred & \small the rigor \\
\hline
\multirow{3}{*}{i mean to}
& \small and be made & \small i seen to \\
& \small be a banner & \small i mean you \\
& \small no less than & \small we mean to \\
\Xhline{1pt}
\end{tabular}
\label{table:libriphrase}
\vspace{-6pt}
\end{table}

\noindent\textbf{Datasets.} 
For training and evaluation, we built a new phrase-unit corpus from the LibriSpeech dataset~\cite{panayotov2015librispeech}.
We call it as \textit{LibriPhrase}~\footnote{\href{https://github.com/gusrud1103/LibriPhrase.git}{https://github.com/gusrud1103/LibriPhrase.git}}, and its creation process is motivated by the data processing in~\cite{lugosch2018donut}.
The training set of LibriPhrase is generated from the \textit{train-clean-100} and \textit{train-clean-360} datasets of LibriSpeech.
LibriSpeech does not provide alignment annotations for audio and text pairs, so we use the Montreal Forced Aligner~\cite{mcauliffe2017montreal} to determine word-level time stamps in the audio. Then, we split full-length utterances into shorter phrases between 1 to 4 words long.
The training set contains 200k phrases of each length, for a total of 800k phrases.

For environmental robustness, we add babble noises of the MS-SNSD dataset~\cite{reddy2019scalable} to audio samples in LibriPhrase during the training, where the SNR is randomly set between 5 to 15dB.
For the evaluation, we generate 4,391~/~2,605~/~467~/~56 episodes of each length consisting of positive and negative audio phrases for enrolled keywords~(anchor) using the \textit{train-others-500} subset. 
Each episode contains 3 positive and 3 negative audio-text pairs.
There are two types of negative sets: easy negatives ($\mathbf{LP_{E}}$) and hard negatives ($\mathbf{LP_{H}}$).
We compute the Levenshtein distances~\cite{levenshtein1966binary} between the anchors and negatives to determine whether a negative sample is easy or hard. \tabref{table:libriphrase} shows examples of easy and hard negatives in LibriPhrase.

In the evaluation phase, we use two more datasets, as well as LibriPhrase.
From the Google Speech Commands V1 dataset ($\mathbf{G}$)~\cite{warden2018speech}, we choose 10 different short words, following the evaluation data setup in~\cite{choi2019temporal}. 
The Qualcomm Keyword Speech dataset ($\mathbf{Q}$)~\cite{kim2019query} consists of the following 4 commands: `Hey android', `Hey snapdragon', `Hi galaxy', and `Hi lumina'.

\begin{table*}
\centering
\caption{Experimental results of UDKWS methods in various datasets. \textbf{G}: Google Commands V1, \textbf{Q}: Qualcomm Keyword Speech dataset, \textbf{LP}$_{\textbf{E}}$: LibriPhrase-Easy, \textbf{LP}$_{\textbf{H}}$: LibriPhrase-Hard. (I): single-modal approach, (II): cross-modal approach}
\vspace{-2pt}
{\normalsize
\begin{tabular}{C{3cm}|C{1cm}|C{1cm}|C{1cm}|C{1cm}|C{1cm}|C{1cm}|C{1cm}|C{1cm}}
\Xhline{1pt}
\multirow{2}{*}{Method} & \multicolumn{4}{c|}{EER~(\%)} & \multicolumn{4}{c}{AUC~(\%)} \\
\cline{2-9}
 & \bf G & \bf Q & \bf LP$_{\bf E}$ & \bf LP$_{\bf H}$ & \bf G & \bf Q & \bf LP$_{\bf E}$ & \bf LP$_{\bf H}$\\
\hline
(I) CTC~\cite{lugosch2018donut} & 31.65 & 18.23 & 14.67 & 35.22 & 66.36 & 89.69 & 92.29 & 69.58\\
(I) Attention~\cite{huang2021query} & \bf 14.75 & 49.13 & 28.74 & 41.95 & \bf 92.09 & 50.13 & 78.74 & 62.65\\
(II) Triplet~\cite{sacchi2019open} & 35.60 & 38.72 & 32.75 & 44.36 & 71.48 & 66.44 & 63.53 & 54.88\\
{\bf (II) Proposed} & 27.25 & \bf 12.15 & \bf 8.42 & \bf 32.90 & 81.06 & \bf 94.51 & \bf 96.70 & \bf 73.58\\
\Xhline{1pt}
\end{tabular}}
\vspace{-4pt}
\label{table:comparison}
\end{table*}

\vspace{-2pt}
\begin{table}
\centering
\caption{EER(\%) results for the ablation study}
\vspace{-4pt}
{
\begin{tabular}{c|c|c|c|c|c|c}
\Xhline{1pt}
    $\mathcal{L}_{D}$ & $\mathcal{L}_{DN}$ & $\mathcal{L}_{MM}$ & \bf G & \bf Q & \bf LP$_{\bf E}$ & \bf LP$_{\bf H}$ \\
\hline
\hline
\small \checkmark & \small - & \small - & \small 27.75 & \small 13.54 & \small  \bf 8.20 & \small 33.80 \\
\small \checkmark & \small \checkmark  & \small - & \small 27.65 & \small 13.02 & \small  8.37 & \small \bf 32.63 \\
\small \checkmark & \small \checkmark & \small \checkmark & \small \bf 27.25 & \small \bf 12.15 & \small 8.42 & \small 32.90\\
\Xhline{1pt}
\end{tabular}
}
\label{table:ablation}
\end{table}

\begin{figure}[t]
\begin{minipage}[t]{.45\linewidth}
  \centering
  \centerline{\includegraphics[width=\columnwidth]{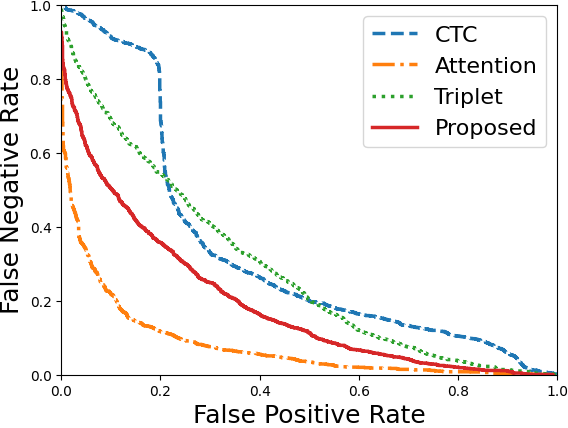}}
 \vspace{-0.05cm}
  \centerline{(a)}
  \centerline{\includegraphics[width=\columnwidth]{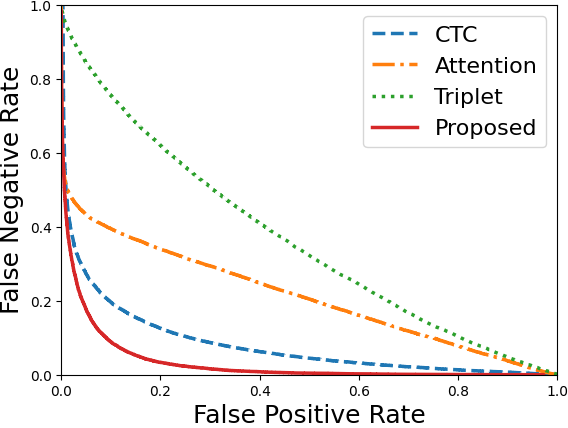}}
 \vspace{-0.05cm}
  \centerline{(c)}
\end{minipage}
\hfill
\begin{minipage}[t]{0.45\linewidth}
  \centering
    \centerline{\includegraphics[width=\columnwidth]{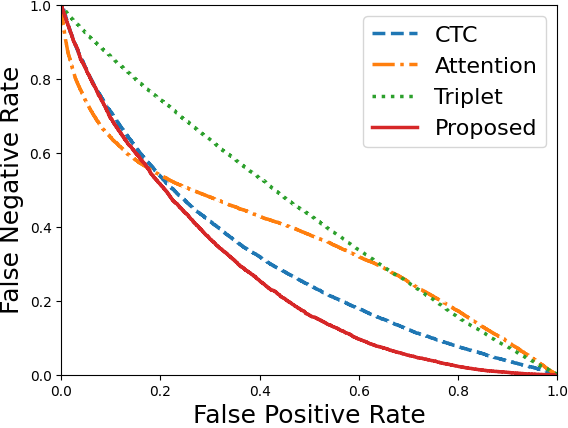}}
 \vspace{-0.05cm}
  \centerline{(b)}
  \centerline{\includegraphics[width=\columnwidth]{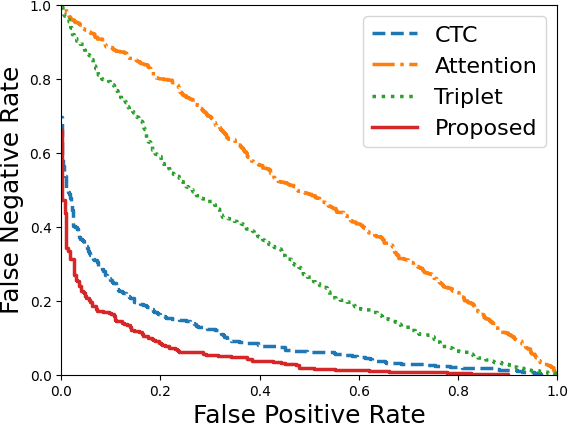}}
 \vspace{-0.05cm}
  \centerline{(d)}
\end{minipage}
\vspace{-4pt}
\caption{DET curves for (a) Google speech commands V1, (b) Qualcomm, (c) LibriPhrase-easy set, (d) LibriPhrase-hard set}
\label{fig:det_curve}
 \vspace{-10pt}
\end{figure}

\begin{figure}[t]
\footnotesize
 \begin{minipage}[b]{.21\linewidth}
  \centering
    \centerline{Text: "i mean to"}
     \vspace{2pt}
  \centerline{\includegraphics[width=\columnwidth]{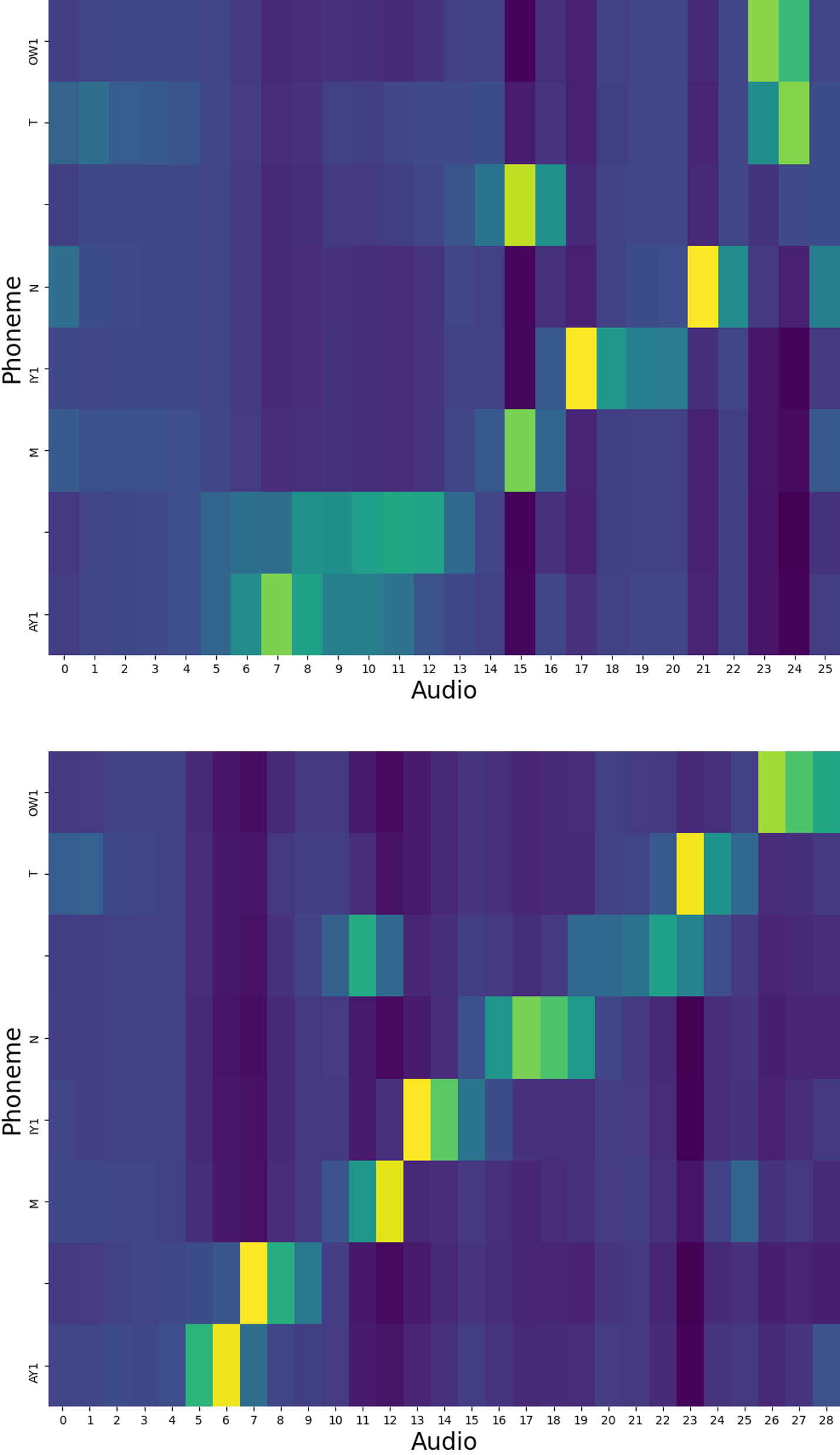}} 
\centerline{(a) full matching}
\end{minipage}
\hfill
\begin{minipage}[b]{0.21\linewidth}
  \centering
    \centerline{"be a banner"}
         \vspace{2pt}
  \centerline{\includegraphics[width=\columnwidth]{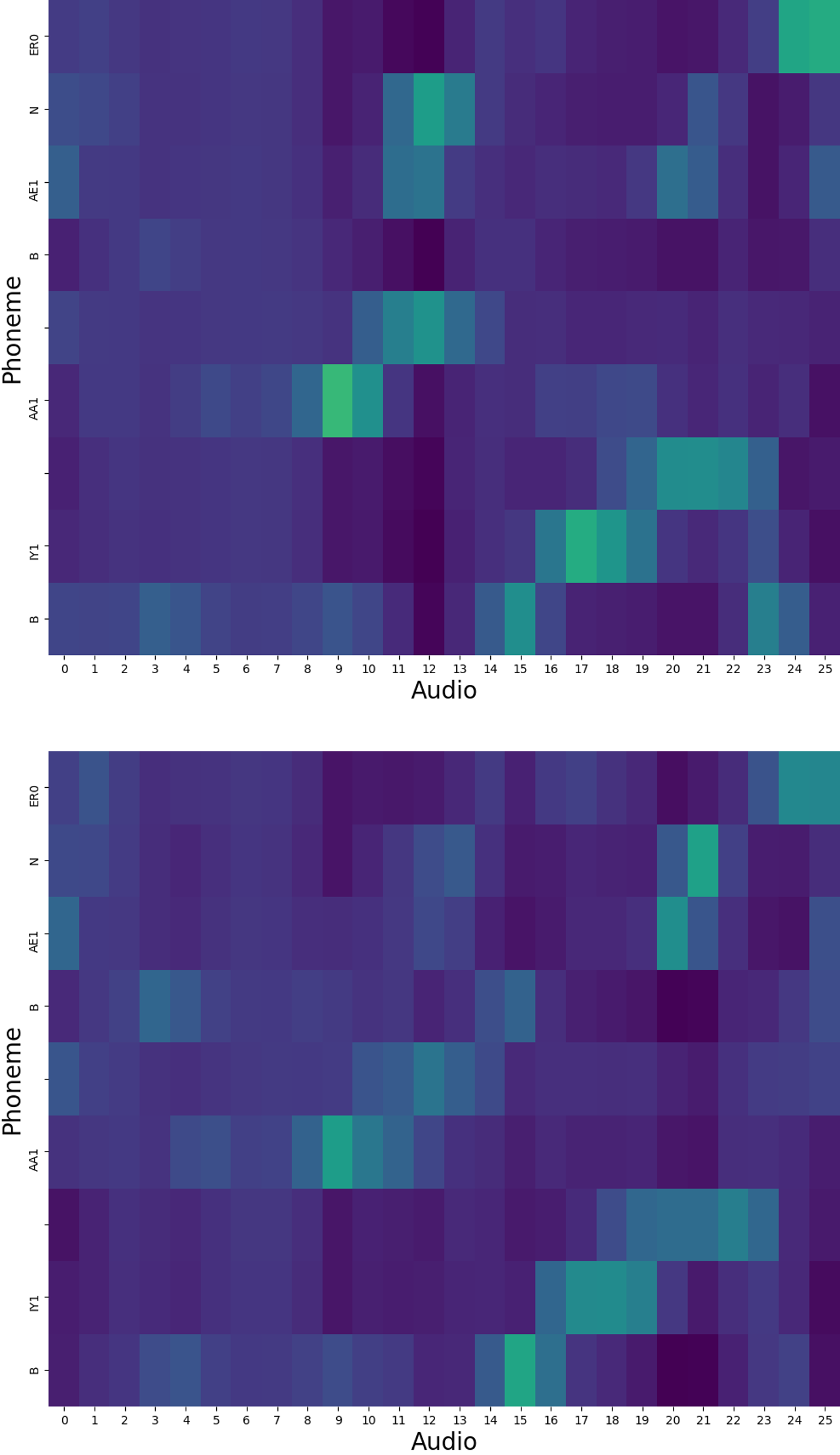}} 
\centerline{(b) non-matching}
\end{minipage}
\hfill
\begin{minipage}[b]{.21\linewidth}
  \centering
  \centerline{"i mean you"}
      \vspace{1pt}
  \centerline{\includegraphics[width=\columnwidth]{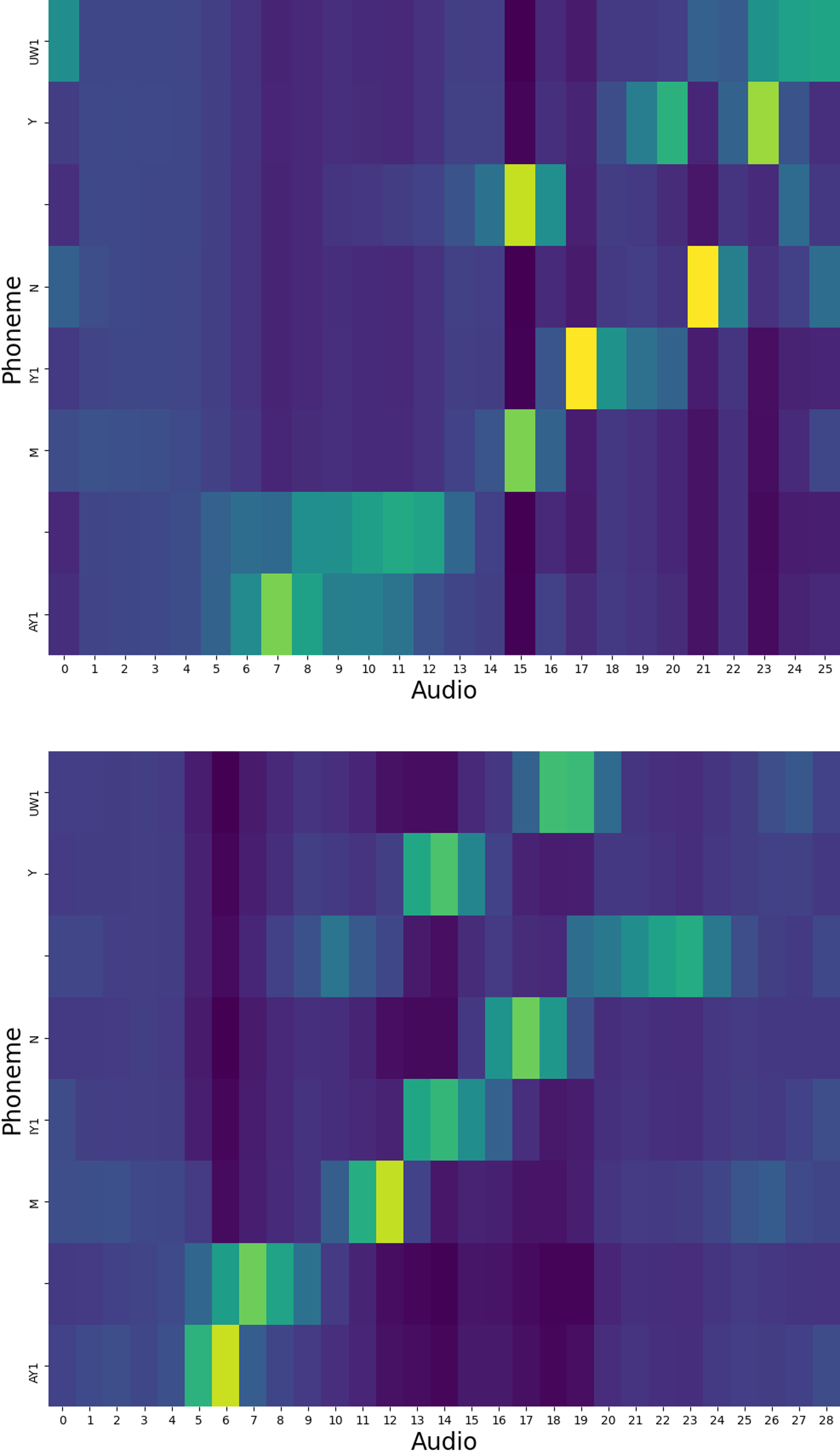}} 
\centerline{(c) partial (front)}
\end{minipage}
\hfill
\begin{minipage}[b]{0.21\linewidth}
  \centering
    \centerline{"we mean to"}
         \vspace{2pt}
  \centerline{\includegraphics[width=\columnwidth]{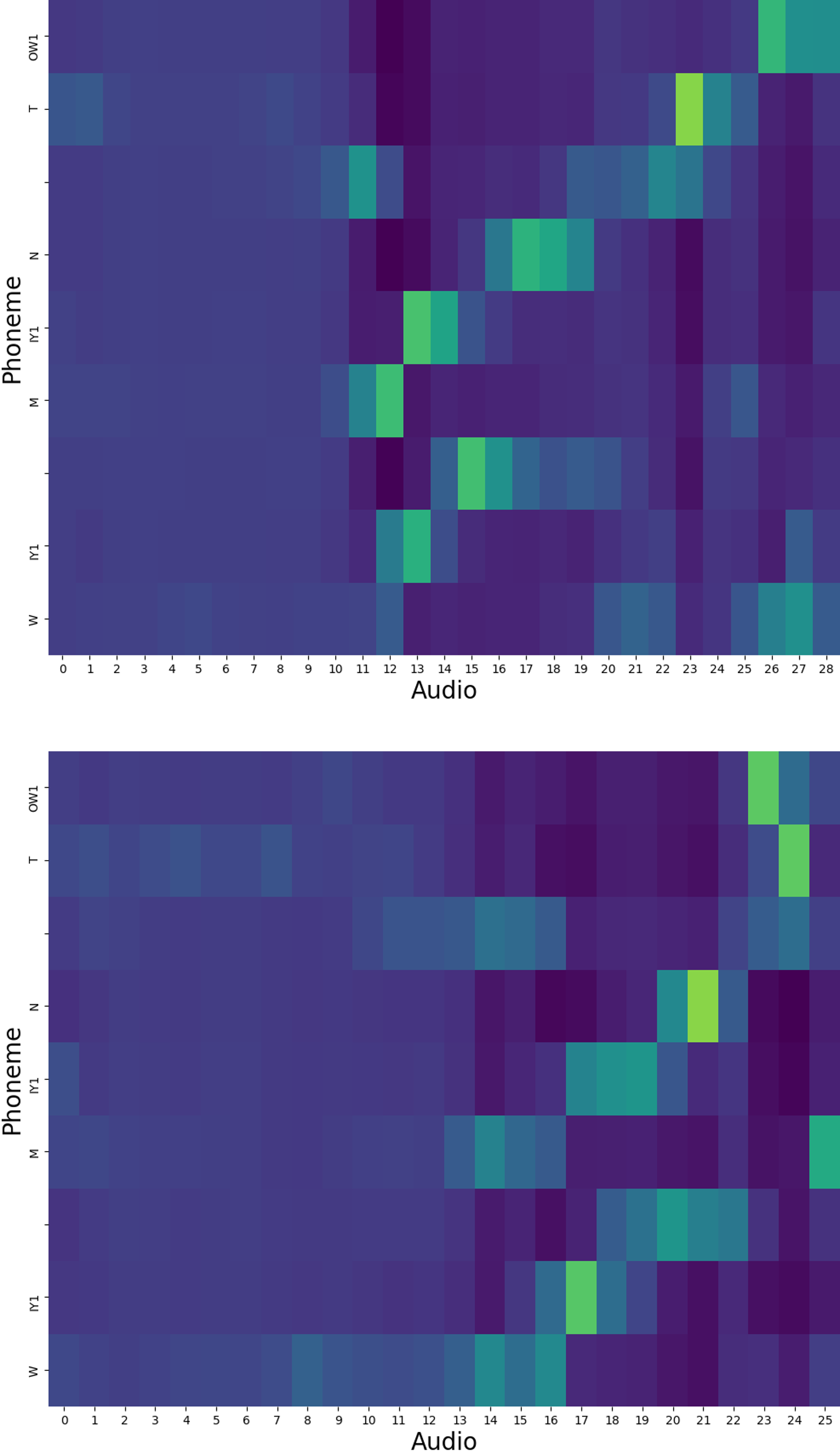}} 
  \centerline{(d) partial (back)}
\end{minipage}
\caption{Qualitative results of ablation study. Visualization of affinity matrices (x-axis: phoneme sequence, y-axis: acoustic frames) for different text keywords when given audio keyword is "i mean to". 1$^{st}$ row: $\mathcal{L}_D$, 2$^{nd}$ row: $\mathcal{L}_D+\mathcal{L}_{DN}+\mathcal{L}_{MM}$.}
\vspace{-5pt}
\label{fig:visualization_of_affinity}
\end{figure}

\input{figure_wordlengths}

\subsection{Experimental results}
We compare performances with conventional approaches trained on the same corpus. We also analyze the impact of each training objective by an ablation study under the same settings.

\vspace{1pt}
\noindent\textbf{Comparison with baselines.} 
All the conventional methods~\cite{lugosch2018donut,huang2021query,sacchi2019open} are re-implemented for the comparison with the proposed model.
Unlike the end-to-end proposed method, they require the enrollment process with the user-defined keyword using 3 audio utterances in the inference stage.
We summarize the results in terms of equal-error-rate~(EER), area under the ROC curve~(AUC), and detection-error-tradeoff~(DET) curve in~\tabref{table:comparison} and~\figref{fig:det_curve}, respectively.

Our proposed model outperforms the conventional methods in the $\mathbf{LP_H}$ and \textbf{Q} dataset.
These datasets have negative phrases that include phonetically similar words to the enrolled keywords.
It is difficult to estimate the phonetic information from the \textbf{Q} dataset because it includes the proper noun at the end of the phrase, \eg, ‘snapdragon’ and ‘lumina’, but our method still outperforms the conventional methods.
On the other hand, the attention-based QbyE method shows powerful performance on \textbf{G} dataset, which consists of frequently used words such as `on',`off',`one' and `two' as keywords.
This method has an advantage when the keyword is included in the training set due to its similarity scoring measurement mechanism.
However, when the keyword is not familiar as in \textbf{Q} and LibriPhrase, it shows degraded performances.
Consequently, our proposed method shows slightly degraded performance compared to the attention-based QbyE method~\cite{huang2021query} that is able to learn the characteristics of the keywords from a classifier in the training stage.
However, except for the \textbf{G} dataset, our proposed method outperforms the baselines on all the datasets in terms of AUC score and EER.

\noindent\textbf{Ablation study.} 
We analyze the effectiveness of each training loss described in~\secref{subsec:loss}.
Table \ref{table:ablation} summarizes the EER results for all the evaluation datasets.
Compared to the results only using detection loss, $\mathcal{L}_{D}$, de-noising loss $\mathcal{L}_{DN}$ improves the overall performance for the \textbf{G} dataset which has environmental distortions. 
However, it has negative impact to the $\mathbf{LP_E}$ dataset that consists of relatively clean audio data. 
In addition, the \matchingloss, $\mathcal{L}_{MM}$, improves overall performance to most of the datasets.
In~\figref{fig:visualization_of_affinity}, we visualize the affinity matrix to show the impact of \matchingloss to various types of matching cases.
It is clear that the \matchingloss helps build more coherent patterns in matching areas, but scattered patterns in non-matching areas.

\noindent\textbf{Analysis on the length of keywords.} 
\figref{fig:length_comparison} depicts the detection performance graphs on the LibriPhrase evaluation dataset. 
It shows the change in performance according to the number of words included in a keyword phrase.
Our proposed method shows robust detection performance regardless of the length of keywords. 
It shows the best scores on most of subsets, while other methods are more vulnerable on shorter keywords.
Especially, the performance of attention-based QbyE method~\cite{huang2021query} significantly degrades when there is a single word in the keyword.

%% file: figure_wordlengths.tex
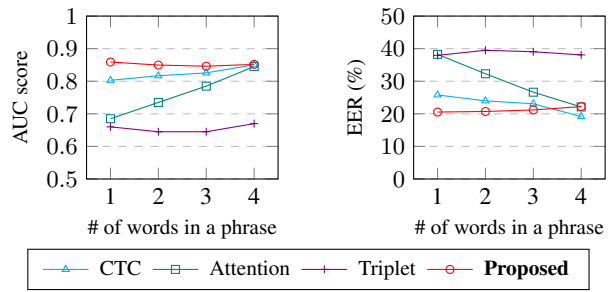
\begin{figure}[t]
    \begin{minipage}[t]{0.45\columnwidth}
        \centering
        \begin{tikzpicture}
        \pgfplotsset{set layers}
        \begin{axis}[
            scale only axis,
            width=0.7*\linewidth,height=0.6\linewidth,
            xlabel={\# of words in a phrase},
            ylabel={AUC score},
            y label style={at={(0.14,0.5)},font=\footnotesize},
            x label style={at={(0.5,0.05)},font=\footnotesize},
            xmin=0.5, xmax=4.5,
            ymin=0.5, ymax=1.0,
            xtick={1, 2, 3, 4},
            xticklabels={1, 2, 3, 4},
            ytick={0.5, 0.6, 0.7, 0.8, 0.9, 1.0},
            ymajorgrids=true,
            grid style=dashed
        ]
            \addplot[
            color=cyan,
            mark=triangle,
            mark size=1.5
            ]
            coordinates
            {(1,0.8028)(2,0.8168)(3,0.8256)(4,0.8505)};\label{auc_ctc}
            \addplot[
            color=teal,
            mark=square,
            mark size=1.5,
            ]
            coordinates
            {(1,0.685)(2,0.735)(3,0.785)(4,0.845)};\label{auc_att}
            \addplot[
            color=violet,
            mark=+,
            mark size=1.5,
            ]
            coordinates
            {(1,0.66)(2,0.645)(3,0.645)(4,0.67)};\label{auc_tri}
            \addplot[
            color=red,
            mark=o,
            mark size=1.5,
            ]
            coordinates
            {(1,0.859)(2,0.8497)(3,0.8462)(4,0.85175)};\label{auc_pro}
        \end{axis}
        \end{tikzpicture}
    \end{minipage}
    \begin{minipage}[t]{0.45\columnwidth}
        \centering
        \begin{tikzpicture}
        \pgfplotsset{set layers}
        \begin{axis}[
            scale only axis,
            width=0.7*\linewidth,height=0.6\linewidth,
            xlabel={\# of words in a phrase},
            ylabel={EER (\%)},
            y label style={at={(0.19,0.5)},font=\footnotesize},
            x label style={at={(0.5,0.05)},font=\footnotesize},
            xmin=0.5, xmax=4.5,
            ymin=0.0, ymax=50,
            xtick={1, 2, 3, 4},
            xticklabels={1, 2, 3, 4},
            ytick={0, 10, 20, 30, 40, 50},
            ymajorgrids=true,
            grid style=dashed
        ]
            \addplot[
            color=cyan,
            mark=triangle,
            mark size=1.5
            ]
            coordinates
            {(1,25.795)(2,23.975)(3,23.09)(4,19.15)};\label{eer_ctc}
            \addplot[
            color=teal,
            mark=square,
            mark size=1.5,
            ]
            coordinates
            {(1,38.255)(2,32.3)(3,26.635)(4,22.125)};\label{eer_att}
            \addplot[
            color=violet,
            mark=+,
            mark size=1.5,
            ]
            coordinates
            {(1,37.945)(2,39.505)(3,39.055)(4,38.095)};\label{eer_tri}
            \addplot[
            color=red,
            mark=o,
            mark size=1.5,
            ]
            coordinates
            {(1,20.545)(2,20.72)(3,21.195)(4,22.22)};\label{eer_pro}
        \end{axis}
        \end{tikzpicture}
    \end{minipage}
    \begin{minipage}[b]{\columnwidth}\fboxsep=0pt\centering
    \fbox{
        \begin{tikzpicture}
        \begin{customlegend}[legend columns=4,legend style={align=left,draw=none,column sep=0.5ex,font=\footnotesize},
                legend entries={CTC,
                                Attention,
                                Triplet,
                                \textbf{Proposed}
                                }]
                \addlegendimage{color=cyan,mark=triangle,mark size=1.5}
                \addlegendimage{color=teal,mark=square,mark size=1.5}
                \addlegendimage{color=violet,mark=+,mark size=1.5}
                \addlegendimage{color=red,mark=o,mark size=1.5}
                \end{customlegend}
        \end{tikzpicture}
    }
    \end{minipage}\hfill
    \vspace{-4pt}
\caption{Evaluation results according to the number of words in a LibriPhrase evaluation set. \textbf{Left}: Area under the ROC curve (AUC), \textbf{Right}: Equal-error-rate (EER)}
\vspace{-5pt}
\label{fig:length_comparison}
\end{figure}

%% file: 4_Conclusion.tex
\vspace{-3pt}
\section{Conclusion}
\label{sec:conclusion}
In this work, we proposed a cross-modal correspondence detector (\proposed), an end-to-end user-defined keyword spotting (UDKWS) system using an agreement between audio and text keyword.
\proposed uses a cross attention mechanism to generate and discriminate meaningful patterns between audio and text embeddings.
We introduced various training strategies for the effective learning, especially using \matchingloss.
We also presented LibriPhrase, a new short-phrase dataset for training the KWS systems in real-world settings.
Experiments showed that \proposed outperformed or showed competitive performance compared to the conventional methods on various datasets.

\vspace{2pt}
\noindent\textbf{Acknowledgements.} 
This research was sponsored by Naver Corporation.